\documentclass[journal]{IEEEtran}
\usepackage{cite}
\usepackage{graphicx}
\usepackage{amsfonts}
\usepackage{amsmath} % assumes amsmath package installed
\usepackage{amssymb}

\newtheorem{thm}{Theorem}
\newtheorem{lem}{Lemma}
\newtheorem{Def}{Definition}
\newtheorem{rem}{Remark}

\newtheorem{assu}{Assumption}

\begin{document}
\title{Stabilizing Quantum States and Automatic Error Correction by Dissipation Control\thanks{This work is supported by the Australian Research Council (DP130101658, FL110100020) and Australian Research Council Centre of Excellence for Quantum Computation and Communication Technology (CE110001027).}}

\author{Yu~Pan,
        Thien~Nguyen% <-this % stops a space
\thanks{Y. Pan is with the Institute of Cyber-Systems and Control, Zhejiang University, Hangzhou 310027, China (e-mail:ypan@zju.edu.cn).

 T. Nguyen is with the Research School of Engineering, Australian National University, Canberra, ACT 0200, Australia (e-mail: thien.nguyen@anu.edu.au).}% <-this % stops a space
}

\maketitle

\begin{abstract}
In this paper an extended scalability condition is proposed to achieve the ground-state stability for a class of multipartite quantum systems which may involve two-body interactions, and an explicit procedure to construct the dissipation control is presented. Moreover, we show that dissipation control can be used for automatic error correction in addition to stabilization. We demonstrate the stabilization and error correction of three-qubit repetition code states using dissipation control.
\end{abstract}


% \fi
\IEEEpeerreviewmaketitle

\section{introduction}
Stabilization of quantum states is central to the scheme of quantum computation. For instance, universal quantum computation \cite{Raussendorf2001} can be realized provided that we can prepare a giant entangled state often called graph state \cite{Hein2004} or cluster state \cite{Nielsen2004}. In order to stabilize the desired quantum states, the system is coupled to an engineered environment, and the dissipative dynamics would drive the system to the target states \cite{Verstraete2009}.

A finite-level quantum system is defined on a Hilbert space $\mathcal H\simeq \mathbb{C}^N$. Denote the space of bounded operator on $\mathcal H$ as $\mathfrak{B}(\mathcal H)$. A quantum state is characterized by a density operator $\rho\in\mathfrak{B}(\mathcal H)$ satisfying $\mbox{trace}(\rho)=1$ and $\rho\geq0$. In many cases, the interaction between the quantum system and the environment is described by a Markov process, and the dynamical equation of the quantum state $\rho_t$ can be written as
\begin{eqnarray}
\dot{\rho}_t&=&\mathcal L(\rho_t)\nonumber\\
&=&-\mbox{i}[H,\rho_t]+\sum_{j=1}^J L_j\rho_t L_j^\dag-\frac{1}{2}L_jL_j^\dag\rho_t-\frac{1}{2}\rho_tL_j^\dag L_j.\nonumber\\\label{rhoMarkov}
\end{eqnarray}
Here $X^\dag$ denotes the adjoint of an operator $X$, and $[X,Y]=XY-YX$. $H\in\mathfrak{B}(\mathcal H)$ is the system Hamiltonian and $\{L_j\in\mathfrak{B}(\mathcal H),j=1,\cdot\cdot\cdot,J\}$ are system operators that characterize the system-environment couplings. Dissipation control is implemented by engineering the system-environment coupling operators.

The stationary states of Eq.~(\ref{rhoMarkov}) have been studied extensively for the purpose of state stabilization \cite{Spohn76,Ticozzi09,Schirmer10,Ticozzi10,Altafini12}. For a multipartite quantum system, the method of using dissipative dynamics to engineer quantum states has been generalized to the notion of {\it dissipatively quasi-locally stabilizable} (DQLS) states \cite{Ticozzi5259,DBLP:journals/qic/TicozziV14}. The theory of DQLS states proposes a systematic approach to determine whether a given multipartite state is asymptotically stabilizable if local dissipation controls can be engineered. Furthermore, if the quantum state satisfies the DQLS condition, the required multipartite system Hamiltonian and the system-environment coupling operators can be constructively derived.

The aforementioned results are based on (\ref{rhoMarkov}). Alternatively, the desired states can be stabilized by studying the evolution of certain operators. Since the expectation of an operator $V\in\mathfrak{B}(\mathcal H)$ at the state $\rho$ is calculated by $\langle V\rangle_\rho=\mbox{trace}(V\rho)$, the evolution of the operator $V(t)$ can be defined via the relation $\langle V(t)\rangle_{\rho_0}=\langle V\rangle_{\rho_t}$. Note that $V=V(0)$. The generator of this Markov process is given by \cite{Breuer07,JG10}
\begin{eqnarray}
\mathcal{G}(V(t))&=&-\mbox{i}[V(t),H(t)]+\mathfrak{L}(V(t))\nonumber\\
&=&-\mbox{i}[V(t),H(t)]+\sum_{j=1}^J L_j^\dagger(t) V(t) L_j(t)\nonumber\\
&&-\frac{1}{2}L_j^\dagger(t) L_j(t) V(t) - \frac{1}{2}V(t)L_j^\dagger(t) L_j(t).\label{eq:Markov_generator}
\end{eqnarray}

A large class of quantum states, including graph states and cluster states (which are DQLS states as well), can be encoded as the ground states of a multipartite operator taking the form $V=\sum_{i=1}^KV_i$ \cite{Perez08,Verstraete2009,Ticozzi5259}. As a result, state stabilization can be achieved by engineering the dissipation such that the system converges to the ground states asymptotically. The merit of this formulation is that the ground-state stability of $V$ can be established by Lyapunov-type operator inequalities. This scenario has been considered before in \cite{Pan2015}, where a scalability condition is used to prove the ground-state stability of $V$ when each $V_i$ is stabilized individually. However the scalability condition proposed in \cite{Pan2015} does not hold for certain applications, especially when $\{V_i\}$ consist of two-body interactions. An illustrative example can be found in Section~\ref{sec4}.

We have two goals in this paper:
\begin{itemize}
\item Based on operator inequalities, derive an extended scalability condition which is applicable to a wider range of applications where the scalability condition from \cite{Pan2015} is not applicable (Section~\ref{sec31}).
\item Prove that the dissipation control is also capable of automatically correcting certain types of errors that occur in the desired quantum states (Section~\ref{sec32}).
\end{itemize}
Note that ground-state stability does not necessarily guarantee that a particular ground state is stable against errors, because the erroneous state may return to a different ground state under the dissipative dynamics. However, we can prove in Section~\ref{sec32} that if certain types of errors occur to one of the ground states, the dissipation control can steer the erroneous state back to the initial ground state exactly, without any measurement or active feedback. In this regard, this result can be considered as an addition to the existing physical literature on {\it automatic quantum error correction} (AQEC) \cite{Barnes00,Kerckhoff2010,Cohen2014,Kapit15,Ippoliti2015}. The definition of AQEC in the context of this paper is given in Sec.~\ref{secr4}.

It is also worth mentioning that the results of this paper are not intended to be used to characterize whether a generic state is DQLS. Instead, the results are intended to deal with a specialized case, where we apply algebraic methods to achieve the stabilization of the states by imposing scalable Lyapunov-type conditions on the operators. If these conditions are satisfied, then the ground states of the system are DQLS and $V$ is frustration-free\cite{Verstraete2009,DBLP:journals/qic/TicozziV14,Pan2015}.

Notations: $Z_V$ is the space spanned by the ground states of $V$. $\sigma_z=\left(\begin{array}{cc}1&0\\0&-1\end{array}
\right),\sigma_x=\left(\begin{array}{cc}0&1\\1&0\end{array}
\right),\sigma_y=\left(\begin{array}{cc}0&-\mbox{i}\\ \mbox{i}&0\end{array}
\right)$ are Pauli operators acting on a two-level system called qubit. Accordingly, $\sigma_{zi},\sigma_{xi},\sigma_{yi}$ are the Pauli operators defined on the $i$-th qubit. The vectorization of a matrix $A$ is denoted as $\mbox{vec}(A)$, which is a column vector obtained by stacking the columns of the $A$ on top of one another. $A^T$ is the transpose of $A$.

\section{preliminaries and assumptions}\label{sec2}
\subsection{Assumptions and definitions}\label{secr1}
The multipartite quantum system considered in this paper is defined on $\mathcal H=\bigotimes_{m=1}^M \mathcal H_m$ which is a tensor product of Hilbert spaces $\{\mathcal H_m\}$ (each $\mathcal H_m$ is associated with a subsystem). We have the following assumption throughout this paper.
\begin{assu}\label{assumption1}
$V$ can be decomposed as $V=\sum_{i=1}^KV_i$, and $V_i$ is defined on a subset of $\{\mathcal H_m\}$. $\{V_i\}$ are orthogonal projections, i.e. $V_i^2=V_i$ and $[V_i,V_j]=0,i\neq j$. Each $V_i$ is associated with a set of dissipation controls $\{L_j\}$. Each $L_j$ allows the decomposition $L_j=U_{i,j}V_i$ with $U_{i,j}$ being a unitary operator.
\end{assu}
\begin{rem}
$\{V_i\}$ can be regarded as quasi-local operators \cite{Ticozzi5259} since they are defined on a subset of $\{\mathcal H_m\}$. Therefore, the stabilizing dynamics in this paper can be considered as specific realizations for the stabilization of DQLS states.

$V_i^2=V_i\geq0$ is a natural assumption that holds for many applications, e.g. the dissipation control of stabilizer states \cite{Perez08,Verstraete2009,Pan2015}. $\{V_i\}$ being commutative is an intuitive assumption which enables $\{V_i\}$ to share common ground states. Physical examples of the decomposition includes the dissipation control of graph states \cite{Verstraete2009}.
\end{rem}

We also make the following assumption.
\begin{assu}\label{assumption2}
The system Hamiltonian can be written as $H=\sum_{i=1}^KH_i$, where each $H_i$ satisfies $H_i=V_i-g_iI$. Here $-g_i$ is the smallest eigenvalue of the Hermitian operator $H_i$.
\end{assu}
\begin{rem}
It is experimentally possible to engineer Hamiltonian on a multipartite quantum system, e.g. \cite{Kienzler2015}.
\end{rem}

As shown in the next section, the two assumptions allow a concise and scalable stability analysis based on the generator (\ref{eq:Markov_generator}). In addition, we have two definitions as follows.
\begin{Def}
$V_i$ is said to be a two-body operator if it can be decomposed as $V_i=X_{m_1}\otimes X_{m_2}$, with $X_{m_1},X_{m_2}$ defined on two Hilbert spaces $\mathcal H_{m_1},\mathcal H_{m_2}$, respectively.
\end{Def}
\begin{Def}
The generator of the evolution of $V_i$ that is induced by a single dissipation control $L_i$ is defined by
\begin{equation}\label{sgen}
\mathcal G(V_i)_{L_i}=-\mbox{i}[V_i,H]+L_i^\dagger V_i L_i-\frac{1}{2}L_i^\dagger L_i V_i - \frac{1}{2}V L_i^\dagger L_i.
\end{equation}
\end{Def}
\begin{rem}
Eq.~(\ref{sgen}) is the generator of $V_i$ controlled by a single coupling operator $L_i$. If $V_i$ is also affected by other coupling operators $\{L_j,j\neq i\}$, then we have $\mathcal G(V_i)=\mathcal G(V_i)_{L_i}+\sum_{j\neq i}L_j^\dagger V_i L_j-\frac{1}{2}L_j^\dagger L_j V_i - \frac{1}{2}V_i L_j^\dagger L_j$.
\end{rem}

\subsection{Previous results}\label{secr2}
We recall one theorem from \cite{JG10}:
\begin{thm}\label{thmp1}
If an operator $X\geq0$ satisfies the following inequality
\begin{equation}\label{eq:es}
\mathcal{G}(X) \leq -cX,\quad c>0,
\end{equation}
then the system will asymptotically converge to $Z_X$.
\end{thm}
\begin{rem}
The algebraic condition (\ref{eq:es}) uses $X=X(0)$, $H=H(0)$ and $\{L_j=L_j(0)\}$. The satisfaction of this condition implies that $\lim_{t\rightarrow\infty}\langle X(t)\rangle=0$. The other algebraic conditions of this paper also use the operators at the initial time. For the details of the Heisenberg-picture stability theory, please refer to \cite{Pan2014}.
\end{rem}

The following theorem can be derived using Theorem~\ref{thmp1}, Assumption~\ref{assumption1} and \ref{assumption2}.
\begin{thm}\cite{Pan2015}
If the following condition
\begin{equation}\label{condition1}
\sum_{j=1}^J(V_iU_{i,j}^\dagger V_iU_{i,j}V_i-V_i)\leq-c_iV_i,\ c_i>0,
\end{equation}
holds, then $\mathcal{G}(V_i)\leq -c_iV_i$ and $V_i$ is asymptotically ground-state stable under the dissipation control of $\{L_j\}$. In particular, we say $\{U_{i,j}\}$ stabilize $V_i$ if Eq.~(\ref{condition1}) holds.
\end{thm}

Now we recall the scalability condition derived in \cite{Pan2015}.
\begin{thm}
Suppose for each $V_i$, there exists $\{L_{i,j}\}$ such that $\mathcal{G}(V_i)\leq -c_iV_i,c_i>0$ holds. The ground-state stability of $V$ can be implied if the intuitive scalability condition
\begin{equation}
\label{eq:Lyapunov_QEC_coupling_condition}
\sum_{i^{'}\neq i}\sum_j\mathcal{G}(V_{i})_{L_{i^{'},j}} \leq 0,
\end{equation}
can be established for all $i$.
\end{thm}

\begin{figure}
\includegraphics[scale=0.4]{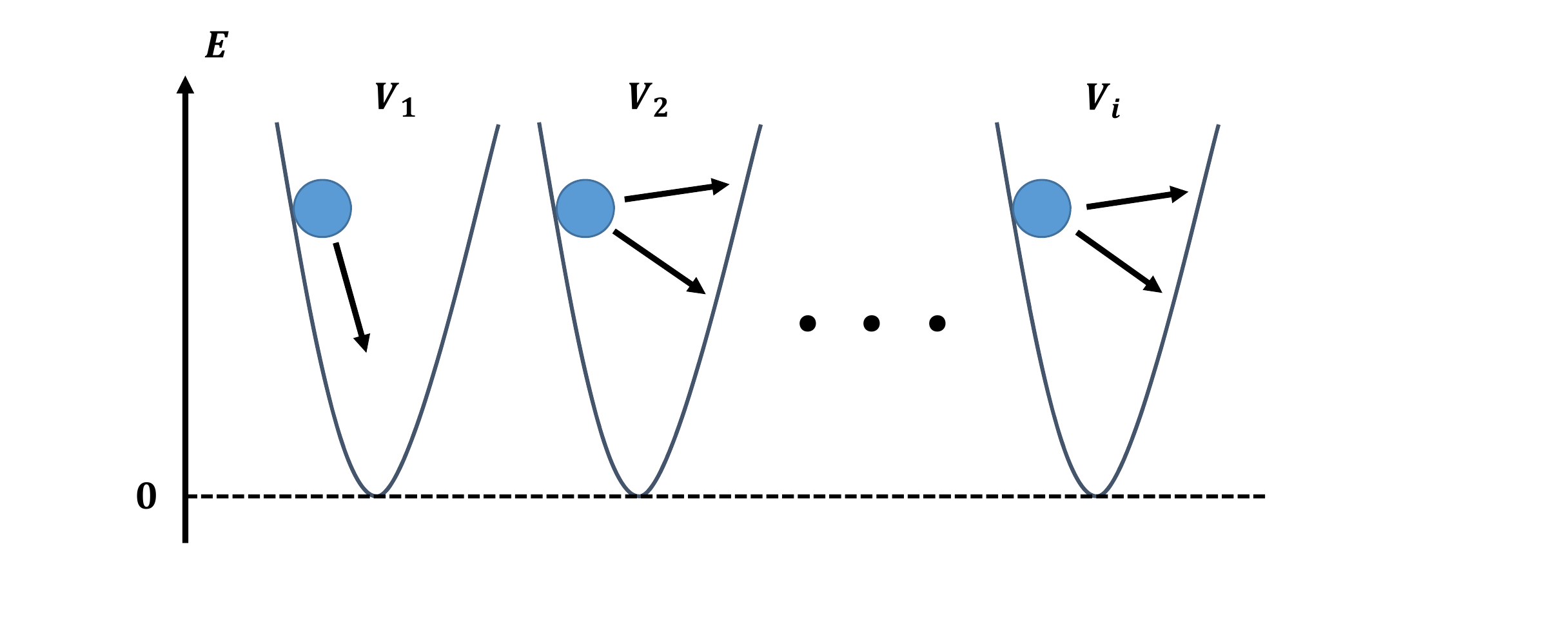}
\caption{Each $V_i$ has $0$ as its lowest energy. The system may be stabilized to the ground state of $V_1$ under a local dissipation control. The local dissipation control is scalable if it maintains or decreases the energy of $\{V_i,i\neq1\}$, i.e. the system is steered towards the ground states of $\{V_i,i\neq1\}$ under the local dissipation control of $V_1$. However for instance, if the local dissipation control of $V_1$ increases the energy of $V_2$, then $\{L_i,i\neq1\}$ must be able to compensate this increase in order to stabilize $V_2$.  }
\label{fig:potential}
\end{figure}

\begin{rem}
$V$ is asymptotically ground-state stable if the local dissipation controls of $\{V_{i^{'}},i^{'}\neq i\}$ do not increase the expectation of $V_i$ (Fig.~\ref{fig:potential}). However, Eq.~(\ref{eq:Lyapunov_QEC_coupling_condition}) is easily violated if $\{V_i\}$ involve two-body interactions. If $V_i$ is a two-body operator, the local control $L_{i,j}$ that stabilizes $V_i$ may not act on $V_{i^{'}},i^{'}\neq i$ trivially. To be more specific, if $V_i=X_{m_1}\otimes X_{m_2}$, then there exists at least one $V_{i^{'}},i^{'}\neq i$ which is defined on $\mathcal H_{m_1}$, otherwise the resulting ground state cannot be entangled in $\mathcal H_{m_1}$. As a result, the local control $U_{i,j}$ defined on $\mathcal H_{m_1}\otimes\mathcal H_{m_2}$ may act nontrivially on this $V_{i^{'}}$.
\end{rem}

The first main result of this paper shows the way to construct the dissipation control for the stabilization of $V$ when $\{V_i\}$ can be locally stabilized but the local dissipation controls do not satisfy the strong scalability condition  Eq.~(\ref{eq:Lyapunov_QEC_coupling_condition}).

\subsection{A new result on stability}\label{secr3}
The following theorem is concerned with the existence of dissipation control.
\begin{thm}\label{thmex}
There always exists a set of unitary operators $\{U_{i,j},j=1,\cdot\cdot\cdot,J\}$ that stabilize $V_i$.
\end{thm}
\begin{IEEEproof}
We provide a constructive method to prove the existence. Since $V_i\geq0$, we can write the spectral decomposition of $V_i$ as $V_i=\sum_nh_n|n\rangle\langle n|,h_n\geq0$, with $\{|n\rangle\}$ being the basis vectors of the Hilbert space. Denote one of the ground states (with eigenvalue $0$) as $|0\rangle\langle0|$. Therefore, if for each $h_j>0$ we choose $U_{i,j}=|j\rangle\langle0|+|0\rangle\langle j|+\sum_{n\neq j,0}|n\rangle\langle n|$, then $V_iU_{i,j}^\dagger V_iU_{i,j}V_i=\sum_{n\neq j}h_n^3|n\rangle\langle n|$. Since $h_n$ equals either $0$ or $1$ for a projector $V_i$, we have $\sum_{n\neq j}h_n^3|n\rangle\langle n|=\sum_{n\neq j}h_n|n\rangle\langle n|$. We can verify that
\begin{equation}
\sum_{j=1}^J(V_iU_{i,j}^\dagger V_iU_{i,j}V_i-V_i)=-\sum_{j=1}^Jh_j|j\rangle\langle j|=-V_i,
\end{equation}
where $\{h_j=1,j=1,\cdot\cdot\cdot,J\}$ is the set of positive eigenvalues. The condition (\ref{condition1}) is satisfied with $c_i=1$.
\end{IEEEproof}
\begin{rem}
Theorem \ref{thmex} provides a specific solution to the exponential stabilization problem. Approaches to choose the set of unitaries have also been discussed in \cite{Verstraete2009}.
\end{rem}

\subsection{Definition of AQEC}\label{secr4}

Next we will introduce the definition of AQEC \cite{Barnes00} that is adapted to the context of this paper.
\begin{Def}\label{prop1}
The set of error operators is denoted as $\{E_a\}$ \cite{Kraus1983}. In the error process, an error may occur with a probability. When the error occurs, the corresponding error operator $E_a$ transforms the initial state $\rho_0\in Z_V$ to the erroneous state as $E_a\rho_0E_a^\dag$. In the correction process that follows the error process, AQEC is defined as the dissipative dynamics that stabilizes the system to an arbitrary initial state $\rho_0\in Z_V$, when an error modelled by an arbitrary error operator from $\{E_a\}$ occurs, i.e.,
\begin{equation}\label{aqecdef}
\lim_{t\rightarrow\infty}(E_a\rho_0E_a^\dag)_t=\rho_0,\quad
\end{equation}
holds for arbitrary $\rho_0\in Z_V$ and $E_a$. In this case, the errors are said to be correctable by the dissipation control.
\end{Def}

For example, $E_a$ can be the bit-flip error operator $\sigma_{x}$ which may flip the state of the qubit as $\sigma_x|0\rangle=|1\rangle$ with certain probability. AQEC will automatically correct the error and return the system state to $|0\rangle$ by dissipation control.

Note that the AQEC condition (\ref{aqecdef}) implies that the system is ground-state stable as every $\rho_0\in Z_V$ is an invariant state. As a result, if no error occurs, the ground states will be stable under the dissipation control. However, ground-state stability does not necessarily imply error correction capability. For instance, suppose the erroneous state is $E_a\rho_0E_a^\dag$. The erroneous state can be automatically steered back to the invariant subspace $Z_V$ if the system is asymptotically ground-state stable, but we cannot guarantee that the dissipation control will restore the system to the initial state $\rho_0$ if the ground states are degenerate, i.e. $Z_V$ is more than one-dimensional. It is also worth mentioning that Definition~\ref{prop1} corresponds to an ideal case that the error process and correction process take place consecutively. Other types of modelling of the two processes can be found in Sec.~\ref{sec4} and \cite{Ippoliti2015}.

\section{main results}\label{sec3}

\subsection{Scalability of dissipation control}\label{sec31}
The following theorem is one of the main results of this paper.
\begin{thm}\label{theorem1}
Suppose $U_i$ stabilizes $V_i$. For each $U_i$, $\{V_j\}$ are separated into two sets, namely, $\{V_n^{(i)}\}$ and $\{V_d^{(i)}\}$. The definitions of the two sets are as follows
\begin{itemize}
\item $[V_n^{(i)},U_i]=0$,
\item $[V_d^{(i)},U_i]\neq0$.
\end{itemize}
A sufficient condition for the asymptotic ground-state stability of $V$ is given by
\begin{eqnarray}
&&V_d^{(i)}U_{i}^\dagger V_d^{(i)}U_{i}V_d^{(i)}\leq V_d^{(i)},\label{condition2}\\
&&\sum_i(\prod_dV_d^{(i)})\geq\lambda V,\ \lambda>0.\label{opcom}
\end{eqnarray}
The dissipation control is constructed as $\{L_i=U_i\prod_dV_d^{(i)}\}$.
\end{thm}
\begin{rem}
By definition, $V_d^{(i)}$ is the operator that satisfies the condition (\ref{condition2}) but does not commute with $U_i$. $U_i$ acts trivially on $\{V_n^{(i)}\}$. In contrast, $U_i$ acts nontrivially on $\{V_d^{(i)}\}$. However, in order for the dissipation controls to be scalable $V_d^{(i)}$ should satisfy the stability condition $\mathcal G(V_d^{(i)})_{U_i}=V_d^{(i)}U_{i}^\dagger V_d^{(i)}U_{i}V_d^{(i)}-V_d^{(i)}\leq0$. Note that $V_i\in V_d^{(i)}$ since $U_i$ stabilizes $V_i$ by assumption.
\end{rem}
\begin{IEEEproof}
By assumption of Theorem~\ref{theorem1}, we have $\{U_i\}$ satisfying Eq. (\ref{condition1}) for each $V_i$. Using $V_i\in V_d^{(i)}$ and the dissipation control $\{L_i=U_i\prod_dV_d^{(i)}\}$ we have
\begin{eqnarray}
\mathcal G(V_i)_{L_i}&=&\prod_dV_d^{(i)}U_{i}^\dagger V_iU_{i}\prod_dV_d^{(i)}-V_i\prod_dV_d^{(i)}\nonumber\\
&=&\prod_dV_d^{(i)}V_iU_{i}^\dagger V_iU_{i}V_i\prod_dV_d^{(i)}-\prod_dV_d^{(i)}\nonumber\\
&\leq&(1-c_i-1)\prod_dV_d^{(i)}=-c_i\prod_dV_d^{(i)},\ c_i>0,\nonumber\\
\end{eqnarray}
where we have made use of the properties $[V_i,V_j]=0$ and $V_i^2=V_i$ from Assumption~\ref{assumption1}. For $j\neq i$, either $[U_j,V_i]=0$ which results in
\begin{eqnarray}
\mathcal G(V_i)_{L_{j}}=\prod_dV_d^{({j})}U_{j}^\dagger V_iU_{j}\prod_dV_d^{({j})}-V_i\prod_dV_d^{({j})}=0,
\end{eqnarray}
or $V_i$ does not commute with $U_{j}$ and satisfies $V_i\in\{V_d^{({j})}\}$. According to (\ref{condition2}), this implies
\begin{equation}\label{invd}
V_iU_{j}^\dagger V_iU_{j}V_i\leq V_i.
\end{equation}
Consequently, we can obtain
\begin{eqnarray}
&&\mathcal G(V_i)_{L_{j}}\nonumber\\
&=&\prod_dV_d^{({j})}U_{j}^\dagger V_iU_{j}\prod_dV_d^{({j})}-V_i\prod_dV_d^{(j)}\nonumber\\
&=&\prod_dV_d^{(j)}V_iU_{j}^\dagger V_iU_{j}V_i\prod_dV_d^{(j)}-V_i\prod_dV_d^{(j)}\nonumber\\
&\leq&0,
\end{eqnarray}
which further leads to
\begin{eqnarray}\label{thm1last}
\mathcal G(\sum_iV_i)&\leq&-\sum_ic_i\prod_dV_d^{(i)}\leq-c_{\min}\sum_i\prod_dV_d^{(i)}\nonumber\\
&\leq&-c_{\min}\lambda\sum_iV_i.
\end{eqnarray}
$c_{\min}$ is the smallest positive number in $\{c_i\}$. Here we have used the condition (\ref{opcom}). By Theorem~\ref{thmp1}, $V=\sum_iV_i$ is asymptotically ground-state stable.
\end{IEEEproof}
\begin{rem}
The strong scalability condition (\ref{eq:Lyapunov_QEC_coupling_condition}) is just a special case of Theorem~\ref{theorem1} with $\{V_d^{({i})}=U_i\}$. Furthermore, Eq.~(\ref{condition2}) implies
\begin{equation}
\mathcal G(V_d^{(i)})_{L_i^{'}=U_iV_d^{(i)}}\leq0.
\end{equation}
Therefore, $\prod_dV_d^{(i)}$ is the product of the operators in $\{V_i\}$ whose energies are not increased by the local controls $\{L_i^{'}=U_iV_d^{(i)}\}$. However, if we implement the dissipation control as $\{L_i^{'}=U_iV_i\}$, the dissipation control may not satisfy the scalability condition  Eq.~(\ref{eq:Lyapunov_QEC_coupling_condition}).

The updated dissipation control $\{L_i=U_i\prod_dV_d^{(i)}\}$ is still in the form of $\{L_i=U_iV_i^{'}\}$ with $V_i^{'}=\prod_dV_d^{(i)}$. Note that $\{V_i^{'}\}$ are still projectors. The reason we preclude the operators $\{V_n^{(i)}\}$ from the updated dissipation controls is that $\prod_dV_d^{(i)}\prod_nV_n^{(i)}=\prod_jV_j$ leads to the same $V_i^{'}$ for each dissipation control and so often results in the violation of sufficient condition (\ref{opcom}) in practical implementation.
\end{rem}

\subsection{Automatic quantum error correction by dissipation control}\label{sec32}
In this section, we derive the condition such that the dissipation control $\{L_i\}$ can automatically correct certain types of errors.

Denote $\{|p\rangle\in\mathcal H\}$ as the complete basis vectors of $Z_V$. Then we can write the basis of the bounded operators on $Z_V$ as $\{|p\rangle\langle q|\}$, where $p$ and $q$ have the same range. An arbitrary density state in $Z_V$ can thus be expanded on the basis as $\rho_0=\sum_{p,q}\alpha_{pq}|p\rangle\langle q|$.

\begin{lem}\label{lemma1}
The dissipation controls $\{L_i\}$ are error-correcting with respect to the error operators $\{E_a\}$ if
\begin{eqnarray}
&&\mathcal L(|p\rangle\langle q|)=0,\label{con51}\\
&&\mathcal L(E_a|p\rangle\langle q|E_a^\dag)\nonumber\\
&&=-\kappa_{pq}E_a|p\rangle\langle q|E_a^\dag+\kappa_{pq}|p\rangle\langle q|,\ \kappa_{pq}>0\label{con5}
\end{eqnarray}
hold for all $p,q$.
\end{lem}

\begin{IEEEproof}
The dynamical equation Eq.~(\ref{rhoMarkov}) can be written as a linear system equation
\begin{equation}\label{linearity}
\dot{\mbox{vec}}({\rho}_t)=A\mbox{vec}(\rho_t),
\end{equation}
using the vectorization of $\rho_t,H,\{L_i\}$ and the relation $\mbox{vec}(B_1B_2B_3)=(B_3^T\otimes B_1)\mbox{vec}(B_2)$. As a result, $A$ is determined by $H$ and $\{L_i\}$. The solution of the above equation is given by
\begin{equation}
\mbox{vec}(\rho_t)=e^{At}\mbox{vec}(\rho_0).
\end{equation}
The condition (\ref{con51}) ensures the invariance of the initial state under the dissipation control, if no error occurs. If an error $E_a$ occurs, the erroneous state $\sum_{p,q}\alpha_{pq}E_a|p\rangle\langle q|E_a^\dag$ needs to be steered back to $\rho_0$. Since we have
\begin{equation}
\mbox{vec}((E_a|p\rangle\langle q|E_a^\dag)_t)=e^{At}\mbox{vec}(E_a|p\rangle\langle q|E_a^\dag),
\end{equation}
and so
\begin{eqnarray}\label{ODE}
&&\dot{\mbox{vec}}((E_a|p\rangle\langle q|E_a^\dag)_t)\nonumber\\
&=&e^{At}A\mbox{vec}(E_a|p\rangle\langle q|E_a^\dag)\nonumber\\
&=&-\kappa_{pq}e^{At}\mbox{vec}(E_a|p\rangle\langle q|E_a^\dag)+\kappa_{pq}\mbox{vec}(|p\rangle\langle q|)\nonumber\\
&=&-\kappa_{pq}\mbox{vec}((E_a|p\rangle\langle q|E_a^\dag)_t)+\kappa_{pq}\mbox{vec}(|p\rangle\langle q|).
\end{eqnarray}
Eq. (\ref{ODE}) is an ordinary first-order differential equation which can be easily integrated to be
\begin{eqnarray}
&&\mbox{vec}((E_a|p\rangle\langle q|E_a^\dag)_t)\nonumber\\
&=&e^{-\kappa_{pq}t}\mbox{vec}(E_a|p\rangle\langle q|E_a^\dag)\nonumber\\
&+&\kappa_{pq}\mbox{vec}(|p\rangle\langle q|)\int_0^te^{-\kappa_{pq}(t-r)}dr\nonumber\\
&=&e^{-\kappa_{pq}t}\mbox{vec}(E_a|p\rangle\langle q|E_a^\dag)+\mbox{vec}(|p\rangle\langle q|)[1-e^{-\kappa_{pq}t}].\nonumber\\
\end{eqnarray}
This proves $(E_a|p\rangle\langle q|E_a^\dag)_t\rightarrow|p\rangle\langle q|$ as $t\rightarrow\infty$. Due to the linearity of the dynamical equation Eq.~(\ref{linearity}), we can deduce that $(\sum_{p,q}\alpha_{pq}E_a|p\rangle\langle q|E_a^\dag)_t\rightarrow\rho_0$ for any error operator $E_a$. The system is restored to the initial state exactly.
\end{IEEEproof}

Eq.~(\ref{con5}) guarantees that every element of the erroneous density state, including the non-diagonal terms which characterize the quantum coherence, can be restored to the initial value. In this sense, Eq.~(\ref{con51})-(\ref{con5}) are more like the definition of quantum error correction, as compared to the sufficient conditions for the errors to be correctable \cite{Knill1997,Ippoliti2015,Barnes00}.

Based on Lemma~\ref{lemma1}, we can prove the following theorem.
\begin{thm}\label{theorem2}
Suppose the dissipation control takes the form $\{L_i=U_iV_i\}$, with $\{V_i\}$ being projectors and $\{U_i\}$ being unitary operators. The sufficient conditions for the set of error operators $\{U_i^\dag\}$ to be correctable are
\begin{eqnarray}
&&V_iU_i^\dag|p\rangle=U_i^\dag|p\rangle,\label{sy1}\\
&&V_jU_i^\dag|p\rangle=0,\ j\neq i,\label{sy2}
\end{eqnarray}
for all $|p\rangle$.
\end{thm}
\begin{rem}
The correctable errors are determined by the available dissipation controls. For example, $U_i^\dag$ can be the bit-flip error operator $\sigma_{x}$. Note that there exist local dissipation controls $\{L_i=\sigma_{xi}V_i\}$ for the stabilization of the graph states \cite{Verstraete2009,Pan2015}. Therefore, in this case the bit-flip errors caused by $\{\sigma_{xi}^\dag\}=\{\sigma_{xi}\}$ are correctable if the sufficient conditions are satisfied.
\end{rem}

\begin{IEEEproof}
It is easy to verify that Eq.~(\ref{con51}) is satisfied since $Z_V$ is an invariant subspace under the dissipation control and $|p\rangle$ is the basis vector of $Z_V$. Recall that $V_i$ is obtained by displacing the system Hamiltonian $H_i$ according to Assumption~\ref{assumption2}. Since the erroneous state vector $U_i^\dag|p\rangle$ is an eigenstate of $V_i$ with the eigenvalue $1$ and an eigenstate of $\{V_j,j\neq i\}$ with eigenvalue $0$, we can conclude that $U_i^\dag|p\rangle$ is also an eigenstate of the system Hamiltonian $H$. Based on this fact, we can remove the unitary dynamics in Eq.~(\ref{con5}) due to
\begin{eqnarray}
&&-\mbox{i}[H,U_i^\dag|p\rangle\langle q|U_i]=-\mbox{i}[V,U_i^\dag|p\rangle\langle q|U_i]\nonumber\\
&=&-\mbox{i}[V_i,U_i^\dag|p\rangle\langle q|U_i]=0,
\end{eqnarray}
and obtain
\begin{eqnarray}\label{lcon}
&&\mathcal L(U_i^\dag|p\rangle\langle q|U_i)\nonumber\\
&=&\sum_{j=1}^KL_jU_i^\dag|p\rangle\langle q|U_i L_j^\dag\nonumber\\
&-&\frac{1}{2}U_i^\dag|p\rangle\langle q|U_i L_j^\dag L_j-\frac{1}{2}L_j^\dag L_jU_i^\dag|p\rangle\langle q|U_i^\dag\nonumber\\
&=&\sum_{j=1}^KU_jV_jU_i^\dag|p\rangle\langle q|U_i V_jU_j^\dag\nonumber\\
&-&\frac{1}{2}U_i^\dag|p\rangle\langle q|U_i V_j-\frac{1}{2}V_jU_i^\dag|p\rangle\langle q|U_i\nonumber\\
&=&|p\rangle\langle q|-U_i^\dag|p\rangle\langle q|U_i.
\end{eqnarray}
Therefore, Eq.~(\ref{con5}) holds with $\kappa_{pq}=1$. By Lemma~\ref{lemma1}, the dissipation control $\{L_i=U_iV_i\}$ can automatically correct errors induced by the error operators $\{U_i^\dag\}$.
\end{IEEEproof}

Eq.~(\ref{sy1})-(\ref{sy2}) imply a one-to-one correspondence between each error operator $U_i^\dag$ to $V_i$. For each error operator $U_i^\dag$, the erroneous state vector $U_i^\dag|p\rangle$ is an eigenvector of $V_i$ with the positive eigenvalue $1$. At the same time, $U_i^\dag|p\rangle$ remains as the ground state of $V_j$ for $j\neq i$. For this reason, $\{V_i\}$ are similar to the error syndrome projectors as proposed in the stabilizer formalism \cite{Gottesman1997}.

\section{Dissipation control of 3-qubit repetition code states}\label{sec4}
Define $|0_L\rangle,|1_L\rangle$ as the basis vectors of a two-level logical qubit. A logical qubit can be encoded using three physical qubits, i.e.
\begin{equation}
\label{eq:three_qubit_codewords}
|0_L\rangle = |000\rangle,\quad|1_L\rangle =|111\rangle,
\end{equation}
where $|0\rangle,|1\rangle$ denote the basis vectors of the physical qubit and $|000\rangle=|0\rangle\otimes|0\rangle\otimes|0\rangle$. This code is proposed to protect quantum information which is stored in span$(|0_L\rangle, |1_L\rangle)$ against single bit-flip noise, where the single bit-flip error operators are given by $\mathcal{E} = \{\sigma_{xi},i=1,2,3\}$. $\sigma_{xi}$ causes a flip of the $i$-th qubit state. The conventional error correction procedure relies on quantum measurement to identify the error and then performs an operation to correct the error accordingly. In this example we demonstrate that the dissipation control can automatically correct the single bit-flip errors as well as stabilize the code states.

The codewords (\ref{eq:three_qubit_codewords}) are common eigenstates of the stabilizers \cite{Gottesman1997}. For the 3-qubit repetition code, the set of stabilizers is given by
\begin{equation}
\mathcal{S} = \{\sigma_{z1}\sigma_{z2}, \sigma_{z2}\sigma_{z3}, \sigma_{z1}\sigma_{z3}\}.
\end{equation}
Based on the stabilizers, we can define $V$ as
\begin{eqnarray}\label{eq:Lyapunov_3qb}
V &=& \frac{1}{2}[(I-\sigma_{z1}\sigma_{z2}) + (I-\sigma_{z2}\sigma_{z3}) + (I-\sigma_{z1}\sigma_{z3})] \nonumber\\
   &=& V_1+V_2+V_3.
\end{eqnarray}
Note that $V_1,V_2,V_3$ are two-body operators. $|0_L\rangle$ and $|1_L\rangle$ are two ground states of $V$. It is easy to verify that $L_1 = \frac{1}{2}\sigma_{x1}(I-\sigma_{z1}\sigma_{z2})$ satisfies
\begin{equation}
\mathcal{G}(V_{1})_{L_1}= L_1^\dagger V_1 L_1 - \frac{1}{2}L_1^\dagger L_1 V_1 - \frac{1}{2}V_1L_1^\dagger L_1= -V_1,
\end{equation}
with $U_1=\sigma_{x1}$. Similarly, we can obtain the local dissipation controls for $V_2,V_3$ as $L_2=\frac{1}{2}\sigma_{x2}(I-\sigma_{z2}\sigma_{z3}),L_3=\frac{1}{2}\sigma_{x3}(I-\sigma_{z1}\sigma_{z3})$, respectively. However, we have
\begin{equation}
\sum_{j=2,3}\mathcal{G}(V_{1})_{L_j} =\sigma_{z1}\sigma_{z2}V_2\nleq 0,
\end{equation}
which violates the scalability condition (\ref{eq:Lyapunov_QEC_coupling_condition}).

Instead, we can construct the correct dissipation control for $V$ using Theorem~\ref{theorem1}. Since $U_1$ stabilizes both $V_1$ and $V_3$, the updated dissipation control would be designed as $L_1 = U_1V_1' = \frac{1}{2}\sigma_{x1}(I-\sigma_{z1}\sigma_{z3})(I-\sigma_{z1}\sigma_{z2})$ with $V_1' = V_1V_3$. $L_2$ and $L_3$ can be derived similarly with $V_2^{'} = V_1V_2$, $U_2=\sigma_{x2}$, $V_3^{'} = V_2V_3$, $U_3=\sigma_{x3}$. Furthermore, we have $V_1^{'} +V_2^{'}+V_3^{'}=V_1V_3+V_1V_2+V_2V_3=\frac{1}{2}(V_1+V_2+V_3)=\frac{1}{2}V$. By Theorem~\ref{theorem1}, $V$ is asymptotically ground-state stable if the 3-qubit system is coupled to a dissipative environment via $L_1,L_2,L_3$.

\begin{figure}
\includegraphics[scale=0.35]{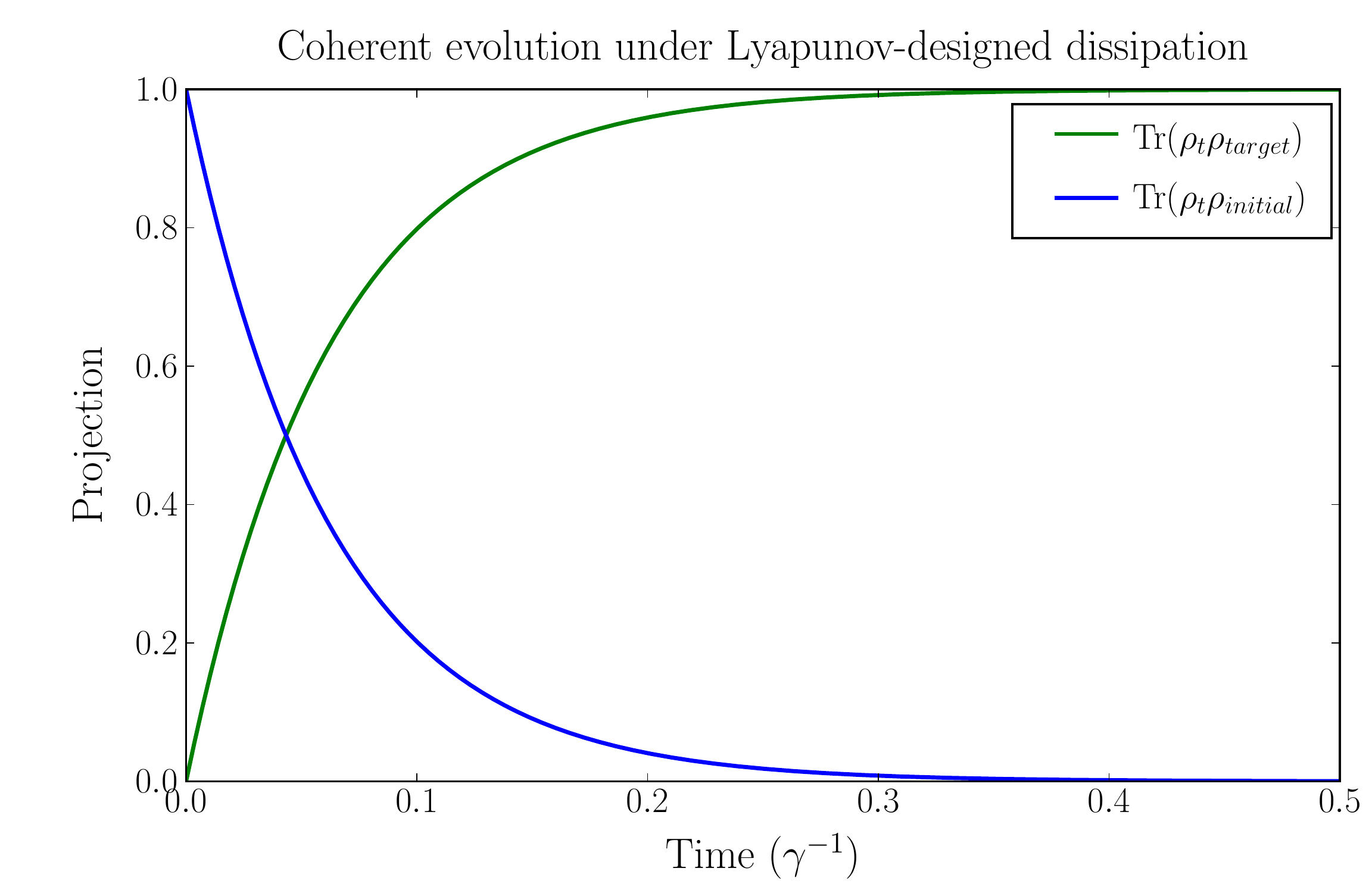}
\caption{Numerical simulation of the state evolution with the coupling operators $L_1^{'},L_2^{'},L_3^{'}$ for 3-qubit repetition code state. Here $L_1^{'},L_2^{'},L_3^{'}$ are renormalized by a coupling strength $\sqrt{\kappa}$, i.e. $L_i^{'}=\sqrt{\kappa}L_i$. The initial state is an error-corrupted state $\rho_{initial} = \frac{|001\rangle+i|110\rangle}{\sqrt{2}}$, which is different from the code state $\rho_0=\rho_{target} = \frac{|000\rangle+i|111\rangle}{\sqrt{2}}$ due to the single bit-flip error $\sigma_{x3}$. The system converges to the target state with unit probability, which means that no coherence is lost during this dissipative process. $\kappa$ and the time scale $\gamma$ are set to $1$ in this example.}
\label{fig:Lyapunov_sim_1}
\end{figure}

Next, we can verify the error correction condition Eq.~(\ref{sy1})-(\ref{sy2}) in Theorem~\ref{theorem2} by
\begin{eqnarray}
&&V_i^{'}\sigma_{xi}|0_L\rangle=\sigma_{xi}|0_L\rangle,\quad V_i^{'}\sigma_{xi}|1_L\rangle=\sigma_{xi}|1_L\rangle, \nonumber\\
&&V_j^{'}\sigma_{xi}|0_L\rangle=0,\quad V_j^{'}\sigma_{xi}|1_L\rangle=0, \ j \neq i.
\end{eqnarray}
As a result, the dissipation control $\{L_i\}$ can automatically correct the errors induced by the set of error operators $\{U_i^\dag\}=\{(\sigma_{xi})^\dag\}=\{\sigma_{xi}\}$, which is exactly the set of the single bit-flip errors $\mathcal{E}$. Fig.~\ref{fig:Lyapunov_sim_1} is the numerical demonstration of the error correction performance of the dissipation control. An erroneous state is shown to be restored to the code state. In particular, the coherence of the initial state is preserved under the dissipation.

\begin{figure}
\includegraphics[scale=0.35]{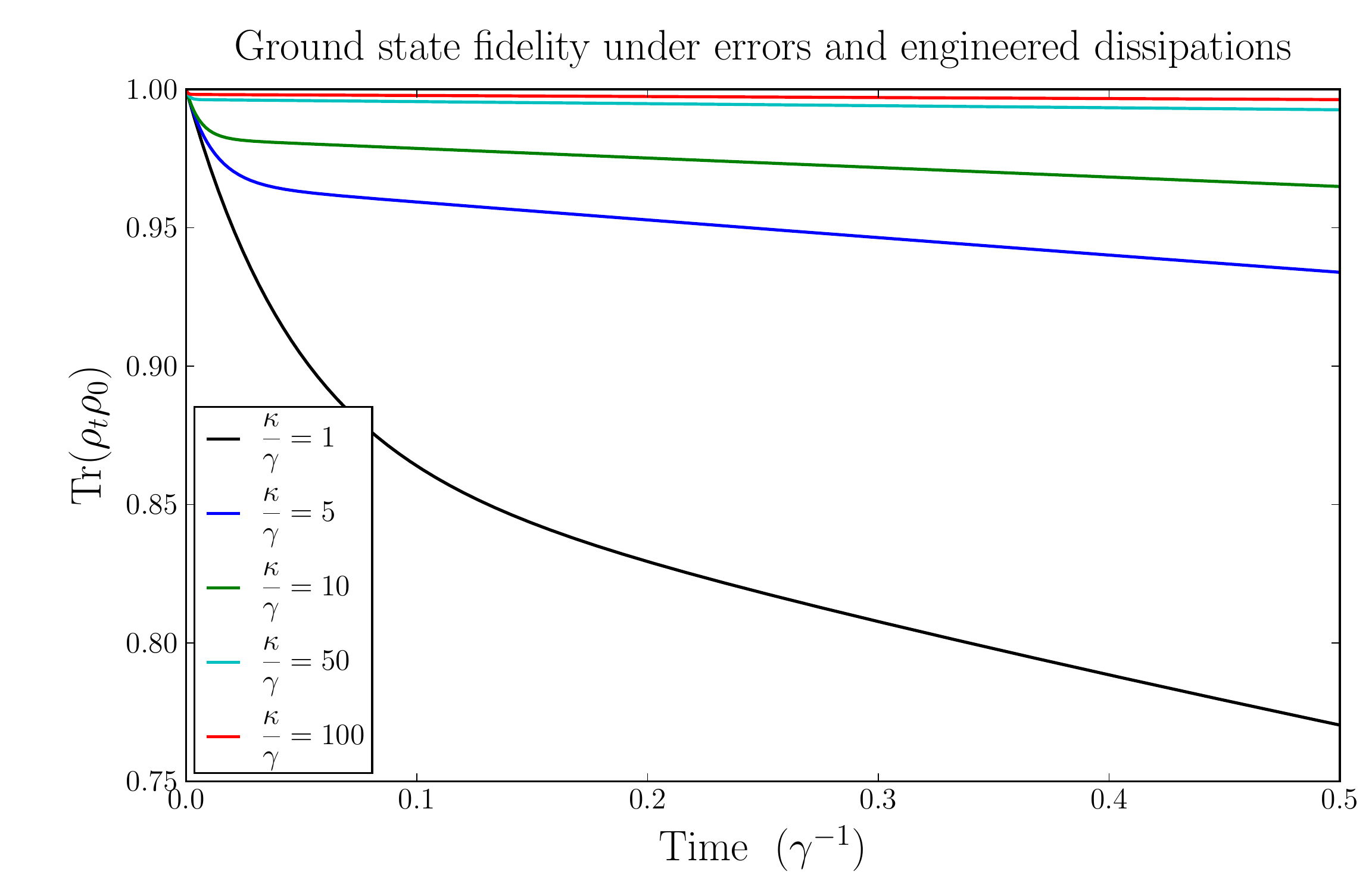}
\caption{Numerical simulation of state evolution when the system is subjected to the dissipative couplings $L_1^{'},L_2^{'},L_3^{'}$ and the noise operators $\{L_i^{noise}=\sigma_{xi},i=1,2,3\}$ simultaneously. $\gamma=1$. The code state is defined by $\rho_0=\frac{|000\rangle+i|111\rangle}{\sqrt{2}}$. Depending on the coupling strength $\kappa$, we achieve different levels of state preservation. In the strong coupling regime $\kappa/\gamma\geq50$, the code state can be continuously preserved against the bit-flip noise.}
\label{fig:Lyapunov_sim_2}
\end{figure}

A more realistic automatic error correction scheme is applying the error correction control in parallel with the noises \cite{Ippoliti2015}. In this case, the system can be modelled as subjected to the noise and the engineered couplings simultaneously. As we have demonstrated in Fig.~\ref{fig:Lyapunov_sim_2}, the derived system-environment couplings for the 3-qubit repetition code states indeed can be used in parallel with the noises. The bit-flip errors are modelled by coupling the system to the environment via the additional coupling operators $\{L_i^{noise}=\sqrt{\gamma}\sigma_{xi},i=1,2,3\}$, with $\gamma$ being the coupling strength. As a result, the system is associated with 6 coupling channels in total. When we increase the strength of the dissipation controls, a nearly perfect code state preservation can be achieved.

\begin{rem} Coherent feedback loop in \cite{Kerckhoff2010} is essentially implementing the environmental couplings adiabatically \cite[Eq. (2)]{Kerckhoff2010}. Physical implementation of dissipation control has also been experimentally demonstrated for superconducting qubit systems, such as in \cite{Shankar2013}. Recently, Cohen \emph{et al.} has demonstrated a scheme which uses the dissipative gadgets to implement automatic quantum error correction \cite{Cohen2014}.
\end{rem}

\section{Conclusion}
Stabilizing ground states is critical for quantum state engineering and quantum computation \cite{Verstraete2009,Pan2015}. We have presented the procedure to construct the dissipation control for the ground-state stabilization of a multipartite quantum system, on which a previous scalability condition may fail to apply if there exist two-body interactions. Moreover, we have investigated the state dynamics of the invariant ground-state space under the dissipation control. We have shown that the dissipation control can automatically correct certain types of errors when these errors occur to the ground states. For these reasons, the dissipation control holds the potential for the protection of noisy qubits, making it a good candidate for the engineering of quantum information by ground-state stabilization \cite{Verstraete2009}.

% or


%\appendices

%\section{}

\section*{Acknowledgment}
The authors would like to thank Matthew James, Valery Ugrinovskii and Michael Hush for their insightful comments.

\ifCLASSOPTIONcaptionsoff
  \newpage
\fi

\end{document}